\documentstyle[12pt,epsf]{article}

\newcommand{\bc}{\begin{center}}
\newcommand{\ec}{\end{center}}
\newcommand{\bd}{\begin{displaymath}}
\newcommand{\ed}{\end{displaymath}}
\newcommand{\be}{\begin{equation}}
\newcommand{\ee}{\end{equation}}
\newcommand{\ba}{\begin{array}}
\newcommand{\ea}{\end{array}}
\newcommand{\bea}{\begin{eqnarray}}
\newcommand{\eea}{\end{eqnarray}}
\newcommand{\bt}{\begin{tabular}}
\newcommand{\et}{\end{tabular}}

\newcommand{\bp}{\begin{picture}}
\newcommand{\ep}{\end{picture}}
\newcommand{\bfi}{\begin{figure}}
\newcommand{\efi}{\end{figure}}

\begin{document}

\title{\huge \bf {String from Veneziano model}}

\author{
H.B.~Nielsen  \footnote{\large\, hbech@nbi.dk} \\[5mm]
\itshape{${}^{1}$ The Niels Bohr Institute, Copenhagen, Denmark}}

\maketitle

\begin{abstract}
This article is about my memories from the
discovery\cite{almost1,almost2,Susskind,Nambu} that the Veneziano
model\cite{Veneziano} described in fact interacting strings. I came
to the understanding that the dual or Veneziano model is really a
model of strings independently of Susskind\cite{Susskind} and
Nambu\cite{Nambu}. A characteristic feature of my approach
\cite{almost1,almost2} was that I used thinking of very high order
``fishnet'' or planar Feynman diagrams as the way of at first
describing the development of the strings. A chain of constituents
lead to planar diagrams dominating when only neighbours on the chain
interact significantly. The article also mentions the works of Ziro
Koba's and mine\cite{nmesons,manifestly,Zeitschrift} about extending
the Veneziano model first to five external particles - as  Bardakci
and Ruegg\cite{BR}, Chan Tsou\cite{Chan} and Goebel and
Sakita\cite{GoebelSakita} also did - and subsequently  to an
arbitrary number $n$ of external mesons.
\end{abstract}

\date{}


\newpage

\thispagestyle{empty}

\section{Introduction}

In a roughly published preprint\cite{almost1,almost2} with the title `` An
almost physical interpretation of the $n$-point Veneziano model''
 I - independently of Nambu \cite{Nambu}  and Susskind\cite{Susskind}
- was one of those who proposed that the dual model or Veneziano
model\cite{Veneziano} was indeed a model describing the scattering
of strings against each other. In fact there was even  an earlier
preprint  version in which the word ``almost'' was not in the title.

It is the purpose of the present contribution to describe my
recollections as to how I came to  understand that the Veneziano
model is a model describing the scattering of strings. The
background that leads to the string idea was the works that developed
dual models first of all to an arbitrary number external particles
together with Ziro Koba\cite{nmesons,manifestly,Zeitschrift}. This
work is described in Section\ref{wwZK}. It would, however, be
natural in such a reminiscence to first describe what for me was
very crucial, namely the seminar by Hector Rubinstein given at the
Niels Bohr Institute. Hector was clearly very excited about the very
great discovery of Veneziano. These earliest memories will be
presented in \ref{Hector}. The line of thinking that ended with the
insight that the Veneziano model describes interacting strings is
described in Section \ref{Vis}. In my case this insight came from using my
pet idea of very high order Feynman diagrams for the purpose of
studying strong interactions and is  presented in section\ref{Sdi}.
I would also like to write about my recollections of the discussions, section
 \ref{dis},
with people at various conferences such as the Lund Conference and
subsequently my correspondence with David Fairlie.

\section{On the Veneziano model} \label{Hector}

Historically Veneziano \cite{Veneziano} started by formulating  a
scattering amplitude for the four mesons
$\pi^+$, $\pi ^0$, $\pi^-$ og $\omega$. This means then that by the
analyticity properties of the Mandelstam representation it should
actually be useful for  the description of the following physical
scattering processes
\begin{itemize}
\item{1}
\begin{equation}
\pi^+ + \pi^o \rightarrow \pi^+ + \omega
\end{equation}
\item{2}
\begin{equation}
\pi^- + \pi^0 \rightarrow \pi^- + \omega
\end{equation}
\item{3}
\begin{equation}
\pi^+ + \pi^- \rightarrow \pi^0 + \omega
\end{equation}
\item{4}
and then the decay process for the unstable $\omega$ to the
three pions
\begin{equation}
\omega \rightarrow \pi^+ + \pi^- + \pi^0
\end{equation}
\end{itemize}

The above list of processes can be extended by taking into account
time reversal symmetry which for example implies the process $\pi^+
+ \omega \rightarrow \pi^+ + \pi^0$ given the process $\pi^+ + \pi^0
\rightarrow \pi^+ + \omega $.


\subsection{(Introduction to the)\\
Mathematical Properties of the
Veneziano Model
Amplitude}
The monumental problem that Gabriele Veneziano succeeded in solving
was to find
a mathematical function of the Mandelstam variables
$s$, $t$
and $u$
(of which only two are needed, since the third one is just a
linear combination of the other two)
with the property that there were no other singularities than the
poles corresponding to the resonances connected with the Regge
poles. Furthermore, in his construction  the assymptotic behavior
was as it should be according to the theory of Regge poles.

Now it must be pointed out that these invented  Regge poles  were
not at first known experimentally although one already knew about
the
$\rho$-meson Regge-trajectory.
Prior to Veneziano´s breakthrough, other physicists had used the
idea of inventing further trajectories (the $\rho$-trajectory was
the only one known experimentally) in order to get a self consistent
amplitude with the combined Regge exchange and Regge resonance
properties. The work preceeding Veneziano must have been an
extremely important source of inspiration for the Veneziano model.
Veneziano then went on to invent infinitely many Regge trajectories.

Let me also explain that the first case for which Veneziano wrote
down the  ``Veneziano model''  had in fact been considered by others
in their efforts to have consistency of the exchange in various
channels. However Veneziano´s first case choice was a very clever
because the isospin properties of the amplitude ensures that the
amplitude must be totally antisymmetric under permutations of the
four momenta of these three pions. (The pions are isovectors that
must couple to the  $\omega$ meson isosinglet).

Veneziano´s choice is implemented by contracting the (totally
antisymmetric) Levi-Civita symbol $\epsilon_{\mu\nu\tau\sigma}$ with
three of the linearly independent four-vectors - i.e., three out of
the four external four momenta - and the polarization four vector
for the $\omega$-particle which is a spin one meson that has its
polarization described by a four vector
$\epsilon^{\mu}$.
This means that the scattering amplitude must contain a factor
\begin{equation}\label{extern}
\epsilon_{\mu\nu\tau\sigma}\epsilon^{\mu}p^{\nu}_{\pi^+}p^{\tau}_{\pi^0}
p^{\sigma}_{\pi^-}.  \label{factor}
\end{equation}

It is this unavoidable factor that we seek to multiply by a function
of the Mandelstam variables $s$, $t$, and $u$. It is upon this
function that analyticity properties are imposed in such a way as to
ensure having only the poles connected with the Regge trajectories
as speculated upon in the Veneziano model. It was the discovery of
this non-trivial function of the Mandelstam variables multiplying
(\ref{factor}) - i.e., the surprisingly simple Veneziano model
function (given in fact by the Euler Beta function) - that was the
great progress made in this Veneziano model.


We may regard this idea of considering a process in which there was
only the possibility for just one factor (\ref{factor}) as just a
detail that was used for getting a simple problem with only one
amplitude as a function of the Mandelstam variables. One could
alternatively have chosen to consider simply scalar particles,
without isospin complications scattering on each other. The latter
was the technology used by Koba and me and also by other groups
including Chan Tsou, Bardakci-Ruegg, Sakita,
etc\cite{BR,Chan,GoebelSakita}
 to get a simple problem appropriate
for  looking for a generalization to five and more external
particles of the Veneziano model. But by choosing scalars one
departs from the  real hadron scatterings of the most important
particles. The lowest mass mesons are indeed pseudoscalars and
vector mesons and they have mostly nontrivial isospins.


After the extraction of the factor (\ref{factor}) the remaining
factor in the amplitude should be symmetric under the permutation of
the three external $\pi$-mesons.
This means in reality that Veneziano had to make this remaining
factor be a sum of three terms coming from each other by the
permutation of these $\pi$-mesons.
But the really interesting question was how these three terms looked
one by one.
In fact each of the three terms has only Regge poles in two out of
the three ``channels''
$s$, $t$ and $u$. So for example it could be that one of the terms
had resonances say only in the s- and the t-channel while there were
no resonances corresponding to particles in the u-channel. It might
be good to have in mind that  an arrangement with the assymmetry
that only the t-channel but not the u-channel has resonances would
be impossible if there were in the s-channel only resonances with
odd spin as would at first be phenomenologically suggested by
considering the $\rho$-meson Regge trajectory in the s-channel. That
would namely suggest that there would only be a sign difference when
we permute the t- and the u-channels. However, there is often the
phenomenon of ``exchange-degeneracy'' meaning that there exist
another trajectory with resonances of even spin together with the
one with odd spins. By interference of the two trajectories with
respectively even and odd spins it becomes possible to arrange the
asymmetry between the s- and the u-channels. So for the Veneziano
model it was rather important to have this exchange degeneracy.

The reason I have mentioned these problems of needing to assume
exchange degeneracy and to have the several amplitudes summed up for
symmetry is because I believe that this was one little technical
detail needed to get a relatively simple solution when seeking to
generalize the Veneziano model to an arbitrary number of external
particles.


This gives me then an excuse for recollecting my cand scient thesis
in which I as a main theme worked on studying the self consistency
of three meson vertices described in the quark model. This may have
strongly inspired me to think of the idea of  processes in which the
mesons are assumed to be composed of different pairs of quark and
antiquarks just so arranged as to guarantee that there would only be
resonances of non-excotic mesons in some of the channels: for
example in the $s$- and $t$-channels but not in the $u$-channel as
in the case just described.

\section{Developing dual models with arbitrary number of external
particles with Ziro Koba}\label{wwZK}

 At the time that Hector gave his for  me sensational seminar on the Veneziano
model\cite{Veneziano} I was rather pressed
to finish my cand. scient thesis\cite{mythesis}. Nevertheless I
managed to include a figure about the Veneziano model in the thesis.
However, first after I had finally delivered the thesis and could do
no more about it did I become so relaxed that I could begin to
concentrate on the Veneziano model. The first work I did concerning
the Veneziano model was done together with Koba. We did the natural
generalization to a five point function. The Veneziano model was
only for a four point scattering amplitude.

What was needed was  an integral formula over two independent
variables that should then obey some algebraic relations so as to
insure the wanted pole and Regge pole behavior at the desired places
in the amplitude as a function of the Regge pole variables
$\alpha(s_{ij})$. These $\alpha(s_{ij})$-variables were put into the
exponents of the dependent variables $u_{ij}$. Here I like to stress
a recollection  of a little accident that may in fact have been of
great significance for me in getting into the project in a fruitful
way. At that time I taught as an instructor -  as I also had done
during my ``eternal´´ studies - a third year course in geometry -
especially projective geometry - at the mathematics department. In
that course we had a problem about a pentagon in the projective
plane in which the students and the instructor had to prove a series
of algebraic relations between anharmonic ratios for this pentagon.
Interestingly enough these relations were exceedingly well suited
for getting  the relations between the dependent variables that
could give the right properties for the generalized dual model for
the scattering of five scalar particles analogous to the Veneziano
model four point function. At that time the most important for my
physics studies and research was the relatively regular meetings
with Ziro Koba and some younger people in a study group. At these
meetings we presented ideas as they came to us. Koba was indeed very
good in inspiring the students to be active. I remember also that he
once ``uninvited´´ me to participate in order that an even younger
group of students should feel more free to express themselves
without feeling embarrassed by the presence of slightly more
advanced people.

In generalizing from the five to the n-point case I think that I
first attempted something like the generalization of the pentagon
formulation. There is indeed such a possibility, but it is more
complicated than the simplified version which Koba discovered. It
was simpler to just have the points on a line in the projective
plane, really the Koba-Nielsen variables.

When we talked about this work of the n-point Veneziano model at
CERN, Koba gave the main talk and I was assigned to talking about
the application of the Haar measure technology needed to make sense
of the expression which would otherwise be divergent.

\section{Ideas that arose in developing string theory}
\label{Sdi} The attempts that led to strings from the Veneziano
model were for me based on the idea of treating strong interactions
by very high order Feynman diagrams. It is of course a most natural
thought that, if the coupling constants are strong, the higher the
order of the diagram the more dominant the diagram should be. At
least the dominant diagrams are expected to be of huge order. So I
set about visualizing such diagrams.

If on the other hand the Veneziano described hadronic scatterings,
should the high order Feynman diagrams not give the Veneziano model?

Then I began to investigate whether I could get the limits in which
one gets the Veneziano model integration variables with the correct
asymptotic behavior and poles from considering very high order
diagrams described mainly as some gross shape structures of the
diagrams. It turned out that approximating the diagrams by a
speculative central limit theorem that rendered  everything Gaussian
- when we convolute suffiently many propagators - was equivalent to
taking the propagators as Gaussian functions of the four momenta.
 It did not quite work out as hoped however
unless a little extra assumption was made. This little extra
assumption was that the diagrams should only be the planar ones.
Assuming this is of course easily argued to be equivalent to
assuming that the particles, the propagators of which were used in
the high order diagrams, tend to sit in long chains. They really
formed strings. Or rather I had to assume that they did sit in such
string chains in order that the summation over the diagrams could
lead to the Veneziano model with the summation over the different
diagrams being identified with the integrations in the Veneziano
model integral representation.

One question that I had to treat was how  to evaluate how the
external momenta from the external lines would get guided through
the diagram dominantly, i.e. from which range in the loop
integration space came the dominant contribution. This is analogous
to the conduction of current through a network of resistances.
Concerning this question I had some discussion that lead to
improvements in my  correspondence with David Fairlie.

My thinking was that if we have a very big - in numbers of loops and
vertices - Feynmann diagram, then  there is an integration over a
huge number of loop four momenta and that therefore the external
four momenta, which of course are led through the diagram (formally
in some arbitrarily specified way), will in some sense only give a
small correction relative to the internal loop four momenta. This
leads to the idea that there is some dominant region in the space of
all the loop momenta around which one gets the main contribution to
the diagram. It is only in this sense of thinking about the dominant
or central point in the integration region over the loop momenta
that one can ask about the ``way that the external momenta are led
through the diagram'' in a physical or mathematically sensible way.
If one knows the point in the integration region that is the center
of the dominant region one can then ask: how in this center have the
contributions to the propagator momenta been changed compared say to
the case when the external momenta were zero. The change could then
be called the way the external momenta are led through the diagram.

If one now plays around with the external four momenta the effects
of changes can be expected to be very small compared to the loop
momenta and it is obvious to imagine making a Taylor expansion in
these small external momenta. In the biggest part of the Feynmann
diagram structure you can indeed simply imagine that because there
are so many ways the external momenta could find their way through
the diagram that it is natural to expect that the external four
momenta can be thought of as being led through in a distributed way.
So only a tiny fraction goes through each of the internal
propagators. Thinking in this way a Taylor expansion would indeed be
suggested as a good approximation. So even though it might in
general not be a good approximation to approximate the propagators
by Gaussian functions of the four momenta as was suggested in my
string paper \cite{almost1,almost2}, at least in very high order complicated
diagrams it should somehow be a good approximation anyway because in
the end the Taylor expansion in the small external four momenta is
what counts.

In this way of arguing one might see some similarity to the
argumentation that in later years has been central in my pet theory
of random dynamics: Often you can Taylor expand so that you get
essentially the same result regardless of the true underlying
theory. So perhaps the true theory does not matter at all in the
end.

\section{Review of my string paper}
\label{Vis}
My preprint(s)\cite{almost1} were poorly distributed. There was a first 
very badly
distributed version with the title ``A physical model for the Dual
model''. In the second slightly better distributed version the word
``almost`` appeared in the title ``An almost physical model for the
Dual model'' as a result of discussions with Knud Hansen who kindly
discussed with me. The concept of a string was put in by the
assumption that the dominant very high order Feynamn diagrams were
{\em planar}. In reality I expressed this assumption by calling them
{\em ``fishnet diagrams''} and drew simply diagrams with lots of
squares as if it were a $\lambda \phi^4$ theory. Of course you can
easily argue that if the constituents of an object - in this case a
hadron or rather a meson - sit sequentially along  a chain forming a
string, the only interaction will be between neighbors on this
chain. The Feynman diagrams that are important would be obtained by
first drawing  the series of propagators representing the
propagation of the constituents (particles)  sitting side by side on
the chain and subsequently connecting them by exchanges that
predominantly are between neighbors only.

Next it is suggested to approximate these diagrams by approximating the
propagators by Gaussian expressions,
\begin{equation}
prop(p^2) \approx C*\exp{(\alpha* p^2)}.
\end{equation}
If one imagines after a transformation to have made it effectively
into  an Euclideanized theory and you use the $+---$ metric, there
needs to be a minus sign in front of the ``C`` in the Gaussian. Here
$C$ is of course just a constant as is $\alpha$.

Gaussian loop integrals can be imagined to be calculated exactly.
Even in practice it is not so difficult to get an estimate of the
dependence on the external momenta, which is the important thing. I
had correspondence with David Fairly who provided important
mathematical insights leading to good analogies for this type of
calculation. Really it is analogous to studying electrical networks
with resisters. Then the approximation of the fishnet diagram by a
planar homogeneous conductor - approximating a conducting fishnet -
becomes quite obvious. One of the important points - also with
inspiration from David Fairly - is that in a two-dimensional network
the flow through of the external momenta is calculable even without
knowing the resistance. There is conformal invariance so that you
can conformally deform the conducting surface without it having any
consequences for the result of the diagram. The external
momentum-dependent factor for the diagram being an exponential of a
quadratic expression has though a coefficient proportional to
$\alpha$ (identified with the resistance).

There are divergence problems connected with leading terms in the
external momenta through point attachments to the conducting disc.
But the really important thing is that the variation of the
positions on the edge of the disc - or the conformally deformed disc
- do not matter because of the conformal invariance. The result is
the integrand in the multiple meson scattering amplitude in the
formulation of Koba and me.

The summation over various large planar diagrams should then
hopefully have led to the integration measure to be used in the formula
by Koba and myself, but that part of the derivation - the correct measure -
did not come (convincingly) out of my technology for obtaining the dual
model.

\subsection{Discussions with many people}\label{dis}
During the time that I developed the string interpretation of the
Veneziano model and especially when I was nearly finished I met with
and discussed with many people many of whom I thank in
acknowledgements.  At the Lund Conference - where my official talk
was about the multiparticle Veneziano model of Koba and me - I
talked with several poeple\footnote{In the acknowledgement are
mentioned Z. Koba, Knud Hansen,Jens Bjoernebo, John Detlefsen, J.
Hamilton, Fisbachin, Jens Hoerup ,D.B. Fairlie, M. Chaichan; and for
encourisment: Virasoro Rubinstein and M. Jacob}
 in the lobbies and of course also with colleagues
at the Niels Bohr Institute.

At one meeting in Copenhagen I should have talked about the string theory
but began to tell the story somewhat privately - I think especially
to Sakita - while the audience was more or less present.  So I gave
the talk in this half private way and did not really get to give the
formal talk. Sakita clearly must have learned my account very well
because at the next big conference in Kiev  he gave a special talk
\cite{almost2}
and subsequently made the first well published work about string
theory using my approach with fishnet diagrams, etc. Neither
Susskind´s nor Nambus´ string theory versions were represented in
Kiev and I believe Sakita had never heard about them.

I remember that I first became aware that Susskind was up to
something similar from a note added at the end of an article by him
telling or at least alluding to that he was making a string theory.
Then I immediately got my mother to send a copy of the preprint to
him. I pressed her so much to hurry that she made remarks about it
that some receiver of preprint were especially favorably treated.

%

 Nambu should have talked about the string at a
Sinbi-conference at the Niels Bohr Institute but got stuck in a
desert and could not come to deliver his talk. But he did deliver
his manuscript.

I also remember that I was honored by being asked to  give a talk
about strings on the occasion of the visit of Heisenberg to The
Niels Bohr Institute. There were two talks that day: mine and that
of Heisenberg. I do not think though that I managed to make
Heisenberg extremely enthusiastic about strings.

\section{Loop correction}

Together with David Fairlie \cite{Fairlie:1970tc} I wrote an article
in which we found that also the loop correction to dual models could
be understood in terms of string theory so to speak.

\section{The vortex line string}

The vortex line, I suppose, should really not be counted as part of
the {\em beginning or birth} of string theory since we came to think
of vortex lines - Poul Olesen and I - only after both having already
worked on and known about the string. So we should rather say
that we attempted to make a model behind the string model. The idea
of vortex lines were at that time already known in the
non-relativistic case of superconductivity, but we were not familiar
with this work. I personally believe to have learned about this work
when presenting our work at CERN. I illustrated a soliton by means
of the necklace of the wife of Masud Chaichans   - and became rather
inspired to make a theory behind the string.

\section{Memories from CERN and understanding the 26 and 10 dimensions}

Following our formulation of string theory as described above, the
period during which I was most occupied with string theory was
during a nine month stay at the theory division of CERN. Here there
was indeed a very good group engaged in working on dual models or
Veneziano models. I had especially strong contact with Paul Frampton
with whom I also visited many good restaurants. Paul worked at that
time on his book about dual models. With Lars Brink I published
among other things a paper which I think provided the nicest way of
seeing what was so special about the critical dimension 26 for the
simple string with only the geometrical degrees of freedom and 10
dimensions for the Neveu-Schwarz-Ramond string having further
fermionic degrees of freedom.

The idea of Lars Brinks and mine \cite{Brink:1986ja} for
understanding the critical dimensions started from the by then well
known point that the first excitation of the string had only the
transverse polarizations and therefore was forced to be massless.
But the point of our idea was to calculate the zero-point energy
contribution to the energy of a string that is not excited but in
its ground state. This contribution to the mass squared of the
string state became - after a renormalization of the speed of light,
as we got it in our formulation - for the simple string the number
$d-2$ of true, transverse, degrees of freedom of the position fields
$X^{\mu}$ multiplied by -1/24 of the distance in mass square between
successive bunches of string states. Here we denoted the number of
``transverse modes'' by $d-2$ so that the number of space-time
dimensions would be $d$. The zero point energy actually shifts the
mass squares for all the string states by the same quantity
$-\frac{d-2}{24} *\frac{1}{\alpha'}$ where $\frac{1}{\alpha}$ is the
mass square spacing between successive levels. In order that the
level, once it becomes excited, gets down to having zero mass
(square) as required for consistency, it is necessary that the shift
down  $-\frac{d-2}{24} *\frac{1} {\alpha'}$ be just equal to the
needed shift $-\frac{1}{\alpha'}$. This gives the famous value for
the space time dimension $d=26$.

\section{Acknowledgement}
It is pleasure to thank Don Bennett for having improoved on the English 
of this article and for Bendt Gudmander also commenting and improving.

\end{document}